\documentclass[aps,preprint,epsfig,rotate]{revtex4}
\usepackage{graphicx}
\usepackage{bm}
\usepackage{epsfig}

\input epsf
\begin{document}
\title{Thermodynamics of the electron-positron plasma at very high temperatures}

\author{Alexei M. Frolov}
\email[E--mail address: ]{alex1975frol@gmail.com}


\affiliation{Department of Applied Mathematics \\
       University of Western Ontario, London, Ontario N6H 5B7, Canada}

\date{\today}

\begin{abstract}

Thermodynamic properties of the electron-positron plasma (or gas) at high and very high temperatures are 
investigated. To achieve this goal we have derived a number of analytical formulas for the Fermi-Dirac 
distribution functions (or spectral functions) which can be applied to various Fermi gases in different 
cases. Almost all these formulas are represented in the form of series expansions. The coefficients in 
these expansions are the explicit and relatively simple functions of the $\frac{\mu}{T}$ ratio, where 
$T$ is the temperature and $\mu$ is the chemical potential of this Fermi system. Our new approach works 
very well for high temperature electron-positron plasma, which is in thermal equilibrium with the photon 
gas of annihilation $\gamma-$quanta, and for the model ultra-relativistic gas of fermions, where there 
is no radiation at all.  \\

\noindent 
DOI: 10.13140/RG.2.2.20786.68801 \\
PACS number(s): 05.30.Fk (Fermion systems and electron gas) and 05.70.Ce (Thermodynamic functions and 
equations of state) 

\end{abstract}

\maketitle
\newpage

\section{Introduction} 

In this study we investigate the basic thermodynamic properties of the Fermi gases, which are confined, e.g.,  
by some very strong gravitational field, at high and very high temperatures. Our main goal is to consider a 
gas which is a mixture of electrons and positrons (elementary particles) which is in thermal equilibrium with 
the photon gas of annihilation $\gamma-$quanta. As follows form \cite{Fro1} such a gas (or plasma) always 
arises inside of any matter (or plasma) heated to extremely high temperatures $T \ge$ 150 $keV$. It was shown 
in \cite{Fro1} for temperature $T =$ 200 $keV$ the total number $N_p$ of the newly created positrons (or 
electrons) and/or electron-positron pairs in the volume of a single hydrogen atom $V_H = \frac{4 \pi 
a^{3}_0}{3}$ exceeds 126, while for $T =$ 300 $keV$ this number equals 2336. After 350 - 400 $keV$ the number 
$N_p$ increases very rapidly and it reaches one million positrons and/or electrons in the volume $V_H$, when 
$T \simeq m c^2$ = 511.0 $keV$. At similar conditions we cannot observe only thermal (or Planck) radiation 
from such an overheated volumes of plasma. In reality, instead of such an object (i.e., gas, or plasma) heated 
to extremely high temperatures, an observer will see only an intense flow of annihilation $\gamma-$quanta, 
which also includes many fast electrons and positrons. Briefly, this phenomenon represents the annihilation 
shielding of overheated objects. In other words, at very high temperatures, which are close to the threshold 
temperature for the electromagnetic vacuum $T \simeq m c^2$ = 511.0 $keV$, our traditional optics ends and we 
have to deal with very intensive streams of outgoing annihilation $\gamma-$quanta, which also include fast 
electrons and positrons. 

Such a high-temperature limit for photon optics has never been assumed either in classical, or quantum optics, 
where it was always believed that one could see all details of `objects' (or bodies as they are called in the 
usual thermodynamic language) heated to arbitrary high temperatures (see, e.g., \cite{Planck}, \cite{BornW} 
and references therein). This conclusion was the most important result of our previous analysis of thermal 
sources of annihilation $\gamma$-quanta in our Galaxy \cite{Fro1}, \cite{Panth1}, \cite{Panth2}. It was 
formulated by the following phrase: due to high-temperature instability of the electromagnetic vacuum and 
extremely intense production of positrons, electrons and annihilation $\gamma-$quanta, it is impossible to 
see (directly) any material object, or plasma of relatively large density, which is heated to the temperatures 
above 350 - 400 $keV$. This phenomenon is called the annihilation shielding of overheated objects. Here we want 
to investigate the phenomenon of annihilation shielding even deeper. At the first step our goal is to understand 
the basic thermodynamic properties of such an electron-positron plasmas of relatively large densities.  
   
A detail description of the Fermi-Dirac statistics and general properties of Fermi gases can be found, e.g., in 
\cite{LLSF} - \cite{EHSE}. These books also contain a number of fundamental facts about Fermi gases at different 
temperatures, which are critically important in our current analysis. In this study we shall consider 
multi-fermion systems, i.e., systems which contain large numbers of fermions, or gas of fermions. The chemical 
potential $\mu$ of such a gas can be either positive, or negative, or equal zero. Our goal in this study is to 
derive the new approach which allows one to describe various thermodynamic properties of these Fermi gases. 
Formally, if we know only a very few (usually two) basic thermodynamic potentials (or properties) of any Fermi 
system, then it is always possible to determine all other thermodynamic functions and potentials which are 
important for thermodynamic analysis. In general, the knowledge of the total number of particles (fermions) $N$ 
and the total energy of the system $E$ as the explicit functions of temperature $T$, volume $V$ and chemical 
potential $\mu$ is sufficient to obtain all thermodynamic properties and potentials of the given Fermi system. 
These two $E(V, T, \mu)$ and $N(V, T, \mu)$ functions are given by the following general formulas (see, e.g., 
\cite{LLSF}, \cite{Feynman}):
\begin{eqnarray}
 N = \int_{0}^{\infty} dN_{\varepsilon} = \frac{g V m^{\frac32}}{\sqrt{2} \pi^{2} \hbar^{3}} 
 \; \int_{0}^{\infty} \frac{\varepsilon^{\frac12} d\varepsilon}{\exp\Bigl(\frac{\varepsilon}{T} - 
 \frac{\mu}{T}\Bigr) + 1} \; \; \; \label{NFNP}
\end{eqnarray}
and 
\begin{eqnarray}
 E = \int_{0}^{\infty} \varepsilon dN_{\varepsilon} = \frac{g V m^{\frac32}}{\sqrt{2} \pi^{2} \hbar^{3}} \;
 \int_{0}^{\infty} \frac{\varepsilon^{\frac32} d\varepsilon}{\exp\Bigl(\frac{\varepsilon}{T} - 
 \frac{\mu}{T}\Bigr) + 1} \; \; , \; \; \label{EFNP}
\end{eqnarray}
where $m$ is the rest mass of the elementary fermion, $\hbar$ is the reduced Planck constant (or Dirac 
constant) and $g = 2 s + 1$ is the spin multiplicity factor, where $s$ is the fermion spin which must 
be half-integer (otherwise the Fermi-Dirac statistics cannot be applied). For the both electrons and 
positrons $s = \frac12$ and $m = m_e$, where $m_e$ is the electron mass at rest. By introducing the new 
variable $x = \frac{\varepsilon}{T} - \frac{\mu}{T}$, i.e., $T dx = d\varepsilon$, we reduce these two 
integrals to the following forms 
\begin{eqnarray}
 N = \frac{g V T^{\frac32} m^{\frac32}}{\sqrt{2} \pi^{2} \hbar^{3}} \; \int^{\infty}_{-\frac{\mu}{T}} 
 \frac{\Bigl(x - \frac{\mu}{T}\Bigr)^{\frac12} dx}{\exp(x) + 1} = V T^{\frac52} 
 f_{N}\Bigl(\frac{\mu}{T}\Bigr) \; \; \; \label{NFNP1}
\end{eqnarray}
and 
\begin{eqnarray}
 E = \frac{g V T^{\frac52} m^{\frac32}}{\sqrt{2} \pi^{2} \hbar^{3}} \;
 \int^{\infty}_{-\frac{\mu}{T}} \frac{\Bigl(x - \frac{\mu}{T}\Bigr)^{\frac32} dx}{\exp(x) + 1} 
 =  V T^{\frac52} f_{E}\Bigl(\frac{\mu}{T}\Bigr) \; \; . \; \; \label{EFNP1}
\end{eqnarray}
In this study we want to derive the closed analytical formulas for these two $\frac{1}{V} \; N(T,\mu)$ 
and $\frac{1}{V} \; E(T,\mu)$ functions which depend upon the temperature $T$ and chemical potential 
$\mu$ only, or in other words, upon the temperature $T$ and the ratio $\frac{\mu}{T}$. This can also 
be written in the form $E(T,\mu) = V T^{\frac52} f_{E}\Bigl(\frac{\mu}{T}\Bigr)$ and $N(T,\mu) = V 
T^{\frac32} f_{N}\Bigl(\frac{\mu}{T}\Bigr)$ and our aim below is to derive the closed analytical 
formulas for the unknown $f_{E}\Bigl(\frac{\mu}{T}\Bigr)$ and  $f_{N}\Bigl(\frac{\mu}{T}\Bigr)$ 
functions. 

Note that if for some Fermi gas the closed analytical formulas for the both $N$ and $E$ functions are known, 
then it is easy to determine other thermodynamic properties of this gas. For instance, for the thermodynamic 
potential $\Omega$ one finds $\Omega = - P V = - \frac23 E$ (the equation of state  \cite{XXX}), while for 
the entropy $S$ we have the formula $S = - \Bigl(\frac{\partial \Omega}{\partial T}\Bigr)_{V,\mu}$. From the 
last formula one finds that the ratio $\frac{S}{N}$ (entropy per one particle) is a homogeneous function of 
zero-order, and therefore, we can write $\frac{S}{N} = \Phi\Bigl(\frac{\mu}{T}\Bigr)$. Analogous formulas 
can be obtained for the pressure $P$ of Fermi gases. In reality, there are many other relations between 
thermodynamic functions which are useful in applications to various Fermi gases \cite{LLSF}, but here we 
cannot discuss them. Instead, we have to follow our main direction and derive the closed analytical formulas 
for the unknown $f_{E}\Bigl(\frac{\mu}{T}\Bigr)$ and $f_{N}\Bigl(\frac{\mu}{T}\Bigr)$ functions in various 
Fermi gases considered at different conditions. This problem is analyzed in the following Sections. In 
general, the results of our method substantially depend upon the sign of chemical potential. For Fermi systems 
such a potential can be either positive (in most of the cases), or negative (for some systems). 
    
\section{Fermi systems with negative chemical potentials} 

In this Section we discuss the systems of fermions which have negative (or non-positive) chemical potentials 
$\mu$. In those cases when $\mu < 0$ the general formulas for the $N$ and $E$ functions can be written in 
one-term universal form which is substantially simpler than analogous formulas derived for the Fermi systems 
with $\mu \ge 0$ (see below). Indeed, if $\mu < 0$, then the total energy $E$ of such a Fermi gas are 
determined by the equation Eq.(\ref{EFNP1}) above. The integral $I$ in Eq.(\ref{EFNP1}) is written in the 
form  
\begin{eqnarray}
  I = \int^{+\infty}_{-\frac{\mu}{T}} \frac{\Bigl(x + \frac{\mu}{T}\Bigr)^{\frac32} dx}{\exp(x) 
  + 1} = \int^{+\infty}_{a} \frac{(x - a)^{\frac32} \exp(- x) dx}{1 + \exp(- x)} \; , \; \label{form1}
\end{eqnarray}
where $a = -\frac{\mu}{T} = \frac{|\mu|}{T} (\ge 0)$ is a positive value. Denominator in this formula 
can be represented as a geometric progression  
\begin{eqnarray}
 & &\frac{1}{1 + \exp(- x)} = 1 - \exp(- x) + \exp(-2 x) - \exp(-3 x) + \exp(-4 x) - \exp(-5 x) 
 \nonumber \\
 &+& \ldots + \nonumber (-1)^{n} \exp(-n x) + \ldots = \sum^{\infty}_{n=0} (-1)^{n} \exp(-n x) 
 \; . \; \; \label{exp}  
\end{eqnarray}
This leads us to the following formula
\begin{eqnarray}
 & &\frac{\exp(- x)}{1 + \exp(- x)} = \exp(- x) - \exp(-2 x) + \exp(-3 x) - \exp(-4 x) + \exp(-5 x) \; 
 \nonumber \\
 &-& \exp(-6 x) + (-1)^{n} \exp[-(n+1) x] + \ldots = \sum^{\infty}_{n=0} (-1)^{n} \exp[-(n+1) x] 
 \nonumber \\
 &=& \sum^{\infty}_{n=1} (-1)^{n-1} \exp(-n x) \; . \; \; \label{exp1}  
\end{eqnarray}
These formulas were often used by Feynman and Mayer in their studies of Fermi systems \cite{Feynman}, 
\cite{Mayer}. For Fermi-systems with negative chemical potentials this formula works perfectly, since for 
such systems $x \ge \frac{|\mu|}{T} > 0$ and the area of integration does not include the `trouble' point 
$x = 0$. By using this formula one obtains the following expression for the integral $I$, Eq.(\ref{exp}): 
\begin{eqnarray}
 I &=& \sum^{\infty}_{n=0} (-1)^{n} \int^{+\infty}_{a} (x - a)^{\frac32} \exp[-(n+1) x] dx = 
 \Gamma\Bigl(\frac52\Bigr) \Bigl[\sum^{\infty}_{n=0} (-1)^{n} \; \frac{\exp[-(n+1) a]}{[(n+1) a]^{\frac52}}
 \Bigr] \nonumber \\ 
 &=& \frac{3 \sqrt{\pi}}{4} \; \Bigl(\frac{1}{a^{\frac52}}\Bigr) \; \Bigl[\sum^{\infty}_{n=0} (-1)^{n} 
 \; \frac{\exp[-(n+1) a]}{(n + 1)^{\frac52}} \Bigr] = \frac{3 \sqrt{\pi}}{4} \; 
 \Bigl(\frac{1}{a^{\frac52}}\Bigr) \; \Bigl[\sum^{\infty}_{n=1} (-1)^{n-1} \; \frac{\exp(-n a)}{n^{\frac52}} 
 \Bigr]\; , \; \label{form2}
\end{eqnarray} 
where the function $\Gamma(x)$ is the $gamma-$function, or Euler's integral of the second kind (see, e.g., 
\cite{GR}). Here we have used the second formula from Eq.(3.382) in \cite{GR}. Analogous formula can be 
derived for the integral $J$ which determines the total number of fermions $N$. The explicit formula is  
\begin{eqnarray}
 J &=& \sum^{\infty}_{n=0} (-1)^{n} \int^{+\infty}_{a} (x - a)^{\frac12} \exp[-(n+1) x] dx = 
 \Gamma\Bigl(\frac32\Bigr) \Bigl[ \sum^{\infty}_{n=0} (-1)^{n} \; \frac{\exp[-(n+1) a]}{[(n+1) 
 a]^{\frac32}} \Bigr] \nonumber \\
 &=& \frac{\sqrt{\pi}}{2} \; \Bigl(\frac{1}{a^{\frac32}}\Bigr) \; \Bigl[\sum^{\infty}_{n=0} (-1)^{n} \; 
 \frac{\exp[-(n+1) a]}{(n + 1)^{\frac32}} \Bigr] = \frac{\sqrt{\pi}}{2} \; \Bigl(\frac{1}{a^{\frac32}}\Bigr) 
 \; \Bigl[\sum^{\infty}_{n=1} (-1)^{n-1} \; \frac{\exp(-n a)}{n^{\frac32}} \Bigr] \; . \; \label{form3} 
\end{eqnarray} 

Now, from the formulas Eqs.(\ref{form2}) and (\ref{form3}) one finds the following expression for the energy 
$E$ and the number of particles $N$ of the Fermi gas which has negative chemical potential $\mu$
\begin{eqnarray}
  E = \frac{3 g V T^{\frac52}}{2 \hbar^{3}} \; \Bigr(\frac{m}{2 \pi}\Bigl)^{\frac32} \;  
  \Bigl(\frac{T}{|\mu|}\Bigr)^{\frac52} \; \Bigl[\sum^{\infty}_{n=1} (-1)^{n-1} \; \frac{1}{n^{\frac52}} \; 
  \exp\Bigl(- \frac{n |\mu|}{T}\Bigr) \Bigr] \; \; . \; \; \label{Enrg3}
\end{eqnarray} 
and 
\begin{eqnarray}
  N = \frac{g V T^{\frac32}}{\hbar^{3}} \; \; \Bigr(\frac{m}{2 \pi}\Bigl)^{\frac32} \; 
  \Bigl(\frac{T}{|\mu|}\Bigr)^{\frac32} \; \Bigl[\sum^{\infty}_{n=1} (-1)^{n-1} \; 
  \frac{1}{n^{\frac32}} \; \exp\Bigl(- \frac{n |\mu|}{T}\Bigr) \Bigr] \; \; . \; \; \label{Nnumb3}
\end{eqnarray} 
Other thermodynamic functions of this Fermi gas can be obtained from these two expressions. For instance, 
thermodynamic potential $\Omega (= P V)$ of the electron (or positron) gas is 
\begin{eqnarray}
 \Omega = - \frac{g V T^{\frac52}}{\hbar^{3}} \; \; \Bigr(\frac{m}{2 \pi}\Bigl)^{\frac32} \; 
 \Bigl(\frac{T}{|\mu|}\Bigr)^{\frac52} \; \Bigl[\sum^{\infty}_{n=1} (-1)^{n-1} \; 
 \frac{1}{n^{\frac52}} \; \exp\Bigl(- \frac{n |\mu|}{T}\Bigr) \Bigr] , \; \; \label{Omega3}
\end{eqnarray} 
where the multiplicity factor $g$ equals $2 s + 1 = 2$ and $s$ is the half-integer spin of the single 
fermion, e.g., for the electron and/or positron gases $g = 2$. The formulas, Eqs.(\ref{Enrg3}) - 
(\ref{Omega3}), are needed to determine all thermodynamic properties of the Fermi gases with negative 
chemical potentials. Indeed, as we have mentioned in the Introduction the knowledge of the total number 
of particles (fermions) $N$ and the total energy of the system $E$ as the explicit functions of 
temperature $T$, volume $V$ and chemical potential $\mu$ is sufficient to obtain all essential 
thermodynamic properties and potentials for a given Fermi system.  

\section{Fermi systems with positive chemical potentials} 

Now, let us consider the fermion systems (or fermion gases of elementary particles) which have positive 
(or non-negative) chemical potentials $\mu$. As mentioned above, to determine the basic thermodynamics 
properties of such a Fermi gas we need to obtain some closed analytical expressions for the total number 
of particles (fermions) $N(V, T, \mu)$ of the Fermi gas located in a given volume $V$ and for the total 
energy $E(V, T, \mu)$ of this gas. These functions are determined by the following general formulas 
\begin{eqnarray}
 E = \frac{g V m^{\frac32}}{\sqrt{2} \pi^{2} \hbar^{3}} \; \int_{0}^{\infty} \frac{\varepsilon^{\frac32} 
 d\varepsilon}{\exp\Bigl(\frac{\varepsilon}{T} - \frac{\mu}{T}\Bigr) + 1} = \frac{g V T^{\frac52} 
 m^{\frac32}}{\sqrt{2} \pi^{2} \hbar^{3}} \; \int^{+\infty}_{-\frac{\mu}{T}} \frac{\Bigl(x + 
 \frac{\mu}{T}\Bigr)^{\frac32} dx}{\exp(x) + 1} \; \; , \; \; \label{EFNP}
\end{eqnarray}
and 
\begin{eqnarray}
 N = \frac{g V m^{\frac32}}{\sqrt{2} \pi^{2} \hbar^{3}} \;  \int_{0}^{\infty} \frac{\varepsilon^{\frac12} 
 d\varepsilon}{\exp\Bigl(\frac{\varepsilon}{T} - \frac{\mu}{T}\Bigr) + 1} = \frac{g V T^{\frac32} 
 m^{\frac32}}{\sqrt{2} \pi^{2} \hbar^{3}} \; \int^{+\infty}_{-\frac{\mu}{T}} \frac{\Bigl(x + 
 \frac{\mu}{T}\Bigr)^{\frac12} dx}{\exp(x) + 1} \; \; , \; \; \label{NFNP}
\end{eqnarray}
where the new variable $x = \frac{\varepsilon}{T} - \frac{\mu}{T}$, where $\frac{\mu}{T} \ge 0$, is used. 
Each of the integrals included in these two formulas is represented as a sum of the two following integrals 
\begin{eqnarray}
 I_{q} = \int_{-\frac{\mu}{T}}^{\infty} \frac{\Bigl(x + \frac{\mu}{T}\Bigr)^{q} dx}{\exp(x) + 1} =
 \int_{-\frac{\mu}{T}}^{0} \frac{\Bigl(x + \frac{\mu}{T}\Bigr)^{q} dx}{\exp(x) + 1} +  
 \int_{0}^{\infty} \frac{\Bigl(x + \frac{\mu}{T}\Bigr)^{q} dx}{\exp(x) + 1} = I^{(1)}_{q} + 
 I^{(2)}_{q} \; \; , \; \; \label{Intq} 
\end{eqnarray} 
where $q = \frac32$ and $\frac12$. By introducing the positive parameter $a = \frac{\mu}{T}$ we can write 
these two integrals in the form 
\begin{eqnarray}
 I_{q} = \int_{-a}^{\infty} \frac{(x + a)^{q} dx}{\exp(x) + 1} = \int_{-a}^{0} \frac{(x + a)^{q} 
 dx}{\exp(x) + 1} + \int_{0}^{\infty} \frac{(x + a)^{q} dx}{\exp(x) + 1} = I^{(1)}_{q}(a) + 
 I^{(2)}_{q}(a) \; \; . \; \; \label{Intqa} 
\end{eqnarray} 

The first integral $I^{(1)}_{q}(a)$ in the last formula is transformed as follows  
\begin{eqnarray}
 I^{(1)}_{q}(a) = \int^{0}_{-a} \frac{(x + a)^{q} dx}{\exp(x) + 1} = \int_{0}^{a} \frac{(-y + a)^{q} 
 dy}{1 + \exp(- y)} = \int_{0}^{a} \frac{(a - y)^{q} dy}{1 + \exp(- y)} \; \; , \; \; \label{Int1qa}
\end{eqnarray}
where we have introduced the new variable $y = - x$ ($d y = - d x$). Now, by applying the formula, 
Eq.(\ref{exp}), we reduce this integral to the form 
\begin{eqnarray}
  I^{(1)}_{q}(a) &=& \sum^{\infty}_{n=0} (-1)^{n} \int_{0}^{a} (a - y)^{q} \exp(-n y) dy =  
  \sum^{\infty}_{n=0} (-1)^{n} B(q+1, 1) \; a^{q+1} \; {}_{1}F_{1}(1, q + 2; -a n) \nonumber \\
  &=& \frac{a^{q + 1}}{q + 1} \Bigl[ \sum^{\infty}_{n=0} (-1)^{n} \; {}_{1}F_{1}(1, q + 2; -a n) 
  \Bigr] \; , \; \label{Int2qa} 
\end{eqnarray}
where we have used the first formula Eq.(3.383) from \cite{GR}. Here $B(x, y) = \frac{\Gamma(x) 
\Gamma(y)}{\Gamma(x + y)}$ is the $beta-$function (Euler's integral of the first kind), ${}_1F_{1}(a, 
b; z)$ is the confluent hypergeometric function, while the function $\Gamma(x)$ is the $gamma-$function, 
or Euler's integral of the second kind. Finally, for $q = \frac32$ and $q = \frac12$ one obtains the  
formulas 
\begin{eqnarray}
 I^{(1)}_{\frac32}(a) = \frac{2 a^{\frac52}}{5} \; \Bigl[ \sum^{\infty}_{n=0} (-1)^{n} \; 
 {}_{1}F_{1}(1, \frac72; -a n)\Bigr] \; \; {\rm and} \; \; I^{(1)}_{\frac12}(a) = 
 \frac{2 a^{\frac32}}{3} \; \Bigl[ \sum^{\infty}_{n=0} (-1)^{n} \; {}_{1}F_{1}(1, \frac52; -a n)\Bigr] 
 \; \; \label{Int3qa} 
\end{eqnarray}
for the two integrals which are needed for our present purposes. Note that the first terms in these two 
functions equal $\frac{a^{q+1}}{q + 1}$ (or $\frac{2 a^{\frac52}}{5}$ and $\frac{2 a^{\frac32}}{3}$, 
respectively), since ${}_{1}F_{1}(1, q + 2; 0) = 1$ for any positive $q$. In actual applications in the 
both equations, Eq.(\ref{Int3qa}), we have to use the fact that $a = \frac{\mu}{T}$. 

Now, consider the second integral $I^{(2)}_{q}(a)$ from the formula, Eq.(\ref{Intqa}). By applying the 
formula, Eq.(\ref{exp}), one finds for this integral  
\begin{eqnarray}
 I^{(2)}_{q}(a) &=& \int^{+\infty}_{0} \frac{(x + a)^{q} \exp(- x) dx}{1 + \exp(- x)} = 
 \sum^{\infty}_{n=1} (-1)^{n-1} \int^{+\infty}_{0} (x + a)^{q} \exp(-n x) dx \nonumber \\
 &=& \sum^{\infty}_{n=1} (-1)^{n-1} \frac{\exp(a n)}{n^{q+1}} \; \Gamma(q + 1, a n) 
 \; , \; \label{form2a}
\end{eqnarray}
where we have used the fourth equation from Eq.(3.382) in \cite{GR}  
\begin{eqnarray}
 \int^{+\infty}_{0} (x + \beta)^{\nu} \exp(-\mu x) dx = \frac{1}{\mu^{\nu + 1}} \; \exp(\beta \mu) 
 \; \Gamma(\nu + 1, \beta \mu) \; \; , \; \; \label{form3a}
\end{eqnarray}
where in our case $\nu = q, \beta = a, \mu = n$ and notation $\Gamma(\alpha, x)$ stands for the incomplete 
$\Gamma$-function defined in \cite{GR} by Eqs.(8.354). The sum of these two integrals $I^{(1)}_{q}(a)$ and 
$I^{(2)}_{q}(a)$ is written in the form 
\begin{eqnarray}
  I^{(1)}_{q}(a) + I^{(2)}_{q}(a) = \frac{2 \; a^{\frac52}}{5} + \sum^{\infty}_{n=1} (-1)^{n-1} \; 
  \Bigl[ \Bigl(\frac{2 \; a^{\frac52}}{5}\Bigr) \; {}_1F_{1}\Bigl( 1, \frac72; -a \; n \Bigr) + 
  \frac{\exp(a n)}{n^{\frac52}} \Gamma\Bigl(\frac52, a n\Bigr) \Bigr] \; \; , \; \; \label{form6a}  
\end{eqnarray}
where $a = \frac{\mu}{T} \ge 0$. The energy $E$ of this Fermi gas is 
\begin{eqnarray}
  E = g V T^{\frac52} \; \frac{1}{\sqrt{2} \pi^2} \Bigl(\frac{m}{\hbar^{2}}\Bigr)^{\frac32} 
  \; \; \Bigl\{ \frac{2 \; a^{\frac52}}{5} &+& \sum^{\infty}_{n=1} (-1)^{n-1} \Bigl[ \Bigl(\frac{2 \; 
  a^{\frac52}}{5}\Bigr) \; {}_1F_{1}\Bigl( 1, \frac72; -a \; n \Bigr) \nonumber \\
  &+& \frac{\exp(a n)}{n^{\frac52}} \Gamma\Bigl(\frac52, a n\Bigr) \Bigr]\Bigr\} \; \; . 
  \; \; \label{form7a}  
\end{eqnarray} 
The total number of particles in this gas $N$ is determined analogously and the final result is represented 
by the formula 
\begin{eqnarray}
  N = g V T^{\frac32} \; \frac{1}{\sqrt{2} \pi^2} \Bigl(\frac{m}{\hbar^{2}}\Bigr)^{\frac32} 
  \; \; \Bigl\{ \frac{2 \; a^{\frac32}}{3} &+& \sum^{\infty}_{n=1} (-1)^{n-1} \Bigl[ \Bigl(\frac{2 \; 
  a^{\frac32}}{3}\Bigr) \; {}_1F_{1}\Bigl( 1, \frac52; -a \; n \Bigr) \nonumber \\
  &+& \frac{\exp(a n)}{n^{\frac32}} \Gamma\Bigl(\frac32, a n\Bigr) \Bigr]\Bigr\} \; \; . 
  \; \; \label{form7a}  
\end{eqnarray} 

The formulas derived in this and previous Sections allow one to determine all basic thermodynamics properties
of the Fermi gases which are located at thermal equilibrium at low, normal and relatively high temperatures 
$T$. However, if temperatures become very high, then fermions must be considered as relativistic particles. 
This means that we have to take into account a number of relativistic and QED effects for these particles. 
First of all, we need to re-derive our formulas for the $N$ and $E$ functions by introducing the rest 
energies of the Fermi particles. Second, we have to take care about radiation which always arise in any 
Fermi gas at high and very high temperatures. Indeed, collisions between electrically charged particles 
always produce a breaking radiation, or bremsstrahlung, which increases with temperature as $T^{\frac12}$.    
Furthermore, interaction of high-temperature radiation with fermions, electrons and atomic nuclei can 
accelerate these particles (inverse bremsstrahlung), which will also produce new fermions in numerous `atomic' 
collisions. In general, for temperatures $T \ge$ 150 $keV$ the production of new electrons and positrons 
becomes very intense. For instance, for temperatures $T \approx m c^{2}$ one atomic volume of a single 
hydrogen atom, i.e., $V = \frac{4 \pi}{3} a^{3}_0$, where $a_0$ is the Bohr radius, contains approximately 
one million newly created electron-positron pairs. These two reasons (relativism and radiation) substantially 
complicate derivation of the explicit formulas for actual thermodynamic functions. Formally, we need to 
derive the new formulas for the $E(V, T, \frac{\mu}{T})$ and $N(V, T, \frac{\mu}{T})$ which can be applied 
to the electron-positron gas (or plasma) at very high temperatures. Below, we consider the two cases which 
are of paramount importance in applications: (a) the electron-positron plasma (gas) at very high temperatures 
which is in thermal equilibrium with annihilation radiation, and (b) a model relativistic gas of fermions 
where annihilation radiation is ignored.  

\section{Electron-positron gas (plasma) at high temperatures} 

In this Section we discuss the electron-positron gas at high and very high temperatures. Here we shall 
assume that thermal energies of the both electrons and positrons are comparable with the electron's  
energy at rest, i.e., $k T \simeq m c^2$, where $m$ is the electron mass at rest and $c$ is the velocity 
of light in vacuum and $k$ is the Boltzmann constant. Everywhere below, we shall express all temperatures 
$T$ (or $kT$) in $keV$. In any substance (or matter) of relatively high density which can be hold at such 
high temperatures for some time one will see formation of very large numbers of electron-positron pairs 
$N_{e-p}$ and their annihilation into $\gamma-$quanta. In reality, already for temperatures $T \ge$ 200 
$keV$ the total number of newly formed electron-positron pairs significantly exceeds the total number of 
initial atomic electrons and nuclei, i.e., particles which were originally present in the same volume 
($V$) before heating. It is clear that at such high temperatures we can neglect (to very good accuracy 
which also increases with $T$) by these atomic electrons and nuclei and their contributions in 
thermodynamic functions and potentials. Therefore, the total numbers of newly created electrons $N_e$ and 
positrons $N_p$ must be equal to each other, i.e., $N_e = N_p$. For chemical potentials of these two Fermi 
gases this means $\mu_e = \mu_{p}$. Note also that at such high temperature electron-positron plasma is  
always in thermal equilibrium with the `photon' gas of annihilation $\gamma-$quanta. Statements that a 
pure positron and/or electron plasma can exist at very high temperatures without radiation is an abstraction 
that fundamentally deviates from reality. This means that the sum of chemical potentials of electron and 
positron gases must be equal to the chemical potential of the gas of photons, i.e., $\mu_e + \mu_{p} = 0$. 
From the equations $\mu_e = \mu_{p}$ and $\mu_e + \mu_{p} = 0$, one finds, that in this case $\mu_e = \mu_{p} 
= 0$. 

First, let us evaluate the total numbers of electrons $N_e$ and positrons $N_p$ at these high temperatures. 
As mentioned above the chemical potentials of the both electron and positron gases equal zero identically. 
In addition to this, at such temperatures we cannot neglect by the rest energy of electron and/or positron 
in the Fermi-Dirac spectral function. Taking into account these two factors, we find that these numbers are 
determined by the following formula
\begin{eqnarray}
  N_e = N_p = \frac{V}{\pi^{2} \hbar^{3}} \int^{+\infty}_{0} \frac{p^{2} dp}{\exp\sqrt{\Bigl(\frac{p 
  c}{T}\Bigr)^{2} + \Bigl(\frac{m c^{2}}{T}\Bigr)^{2}} + 1} \; . \; \label{form2a}
\end{eqnarray}
This formula contains the integral with the Fermi-Dirac spectral function defined above in which $\mu = 0$. 
Here we want to derive an analytical expression for this integral and for the numbers of electrons $N_e$ 
and positrons $N_p$, respectively. Let us introduce the new variable $y = \frac{p c}{T}$ in this integral 
and obtain 
\begin{eqnarray}
  N_e = N_p = \frac{V}{\pi^{2}} \Bigl(\frac{T}{\hbar c}\Bigr)^{3} \int^{+\infty}_{0} \frac{y^{2} 
  dy}{\exp(\sqrt{y^{2} + a^{2}}) + 1} \; , \; \label{form2a}
\end{eqnarray}
where $a^{2} = \Bigl(\frac{m c^{2}}{T}\Bigr)^{2} = \frac{1}{\theta^{2}}$ and $\theta$ is the new temperature 
expressed in the $m c^{2}$ units (energy units). This means that the parameter $a = \frac{1}{\theta}$ is 
always positive. 

To obtain the closed analytical formula for the last integral, Eq.(\ref{form2a}), we introduce the new variable 
$x = \sqrt{y^{2} + a^{2}}$ in this integral. It is clear that for this new variable one finds $d x = \frac{y 
dy}{\sqrt{y^{2} + a^{2}}}$, or $y dy = x dx$. This leads to the following expression 
\begin{eqnarray}
  J = \int^{+\infty}_{a} \frac{x (x^{2} - a^{2})^{\frac12} dx}{\exp(x) + 1} = 
  \int^{+\infty}_{a} \frac{x (x^{2} - a^{2})^{\frac32 - 1} dx}{\exp(x) + 1} \; . \; \label{form3a}
\end{eqnarray}
Since the lower limit in this integral is positive, we can apply the formula, Eq.(\ref{exp}). Finally, we can 
write the $J$ integral as the following sum 
\begin{eqnarray}
 J &=& \sum^{\infty}_{n=1} (-1)^{n-1} \int^{+\infty}_{a} x (x^{2} - a^{2})^{\frac32 - 1} \exp(-n x) dx = 
 a^{2} \Bigl[\sum^{\infty}_{n=1} \frac{(-1)^{n-1}}{n} \; K_{2}(n a) \Bigr] \nonumber \\
 &=& a^{2} \Bigl[ K_{2}(a) - \frac12 K_{2}(2 a) + \frac13 K_{2}(3 a) - \frac14 K_{2}(4 a) + \frac15 K_{2}(5 a) 
  + \ldots \Bigr] \; \; , \; \; \label{form4a}
\end{eqnarray} 
where $K_{2}(x)$ is the modified Bessel function of the second order. The $K_{p}(x)$ functions are also called 
the Macdoanld's functions, since H.M. Macdonald studied and introduced these functions in 1899 \cite{Mac} (see, 
also discussion and references in \cite{Watson}). The final formula for the $N_e = N_p$ numbers takes the form 
\begin{eqnarray}
 N_e = N_p = \frac{V}{\pi^{2}} \Bigl(\frac{T}{\hbar c}\Bigr)^{3} \sum^{\infty}_{n=1} \frac{(-1)^{n-1}}{n \; 
 \theta^{2}} K_{2}\Bigl(\frac{n}{\theta}\Bigr) \; \; . \; \; \label{form5a}
\end{eqnarray}
This formula can also be found in our earlier paper \cite{Fro1}. 

Now, in order to complete our analysis of the electron-positron gas (plasma) at very high temperatures we have 
to derive analogous formula for the energy $E$ of this gas. After a few simple transformations we have found 
that the energies of the electron and positron gases at such high temperatures are evaluated by the formula 
\begin{eqnarray}
 E_e = E_p = \frac{V}{\pi^{2} \hbar^{3}} \int^{+\infty}_{0} \frac{c \; \sqrt{p^{2} + m^{2} c^{2}} \; p^{2} 
 dp}{\exp\sqrt{\Bigl(\frac{p c}{T}\Bigr)^{2} + \Bigl(\frac{m c^{2}}{T}\Bigr)^{2}} + 1} = \frac{V T}{\pi^{2}} 
 \Bigl(\frac{T}{\hbar c}\Bigr)^{3} \int^{+\infty}_{a} \frac{x^{2} (x^{2} - a^{2})^{\frac12} dx}{\exp(x) + 1} 
 \; \; . \; \; \label{form6a}
\end{eqnarray}
where $x$ is our variable defined above, while $a = \frac{m c^{2}}{T} = \frac{1}{\theta}$ and $\theta$ is the 
new temperature expressed in the $m c^{2}$ units (see above). The integral in this formula is represented in 
the form 
\begin{eqnarray}   
 I = \int^{+\infty}_{a} \frac{x^{2} (x^{2} - a^{2})^{\frac12} dx}{\exp(x) + 1} = \sum^{\infty}_{n=1} (-1)^{n-1} 
 \int^{+\infty}_{a} x^{2} (x^{2} - a^{2})^{\frac32 - 1} \exp(-n x) dx \; , \; \label{form7a}
\end{eqnarray}
where we have used the formula, Eq.(\ref{exp}). All integrals included in the last formula have essentially 
identical form. They are determined by using the following general formula 
\begin{eqnarray}   
 \int^{+\infty}_{a} x^{2} (x^{2} - a^{2})^{\nu - 1} \exp(-\mu x) dx &=& \frac{2^{\nu - \frac12} a^{\nu + 
 \frac12}}{\sqrt{\pi}} \Gamma(\nu) \Bigl\{ \Bigl(\frac{\nu - \frac12}{\mu^{\nu + \frac12}}\Bigr) \; 
 K_{\nu + \frac12}(a \mu) \nonumber \\
 &+& \frac{a}{2 \mu^{\nu -\frac12}} \Bigl[ K_{\nu - \frac12}(a \mu) + K_{\nu + \frac32}(a \mu) \Bigr] \Bigr\}
 \; , \; \label{form8a}
\end{eqnarray}
where $K_{p}(x)$ functions are the Macdoanld's functions mentioned above. The formula, Eq.(\ref{form8a}), has 
been derived a few years ago by me. In order to reproduce Eq.(\ref{form7a}) in the last formula we have to 
choose $\nu = \frac32$ and $\mu = n$. After a few simplifications one obtains the following final expression
\begin{eqnarray}      
 \int^{+\infty}_{a} x^{2} (x^{2} - a^{2})^{\frac32 - 1} \exp(-n x) dx = a^{2} \Bigl[ \frac{a}{2 n} K_{1}(n a) 
 + \frac{1}{n^{2}} K_{2}(n a) + \frac{a}{2 n} K_{3}(n a) \Bigr] \; . \; \label{form9a}
\end{eqnarray}
With this result the formula, Eq.(\ref{form6a}), for the energy of electron/positron gas takes the form 
\begin{eqnarray}
 E_e = E_p &=& \frac{V}{\pi^{2} \hbar^{3}} \int^{+\infty}_{0} \frac{c \; \sqrt{p^{2} + m^{2} c^{2}} \; p^{2} 
 dp}{\exp\sqrt{\Bigl(\frac{p c}{T}\Bigr)^{2} + \Bigl(\frac{m c^{2}}{T}\Bigr)^{2}} + 1} \nonumber \\
 &=& \frac{V T}{\pi^{2}} \Bigl(\frac{T}{\hbar c}\Bigr)^{3} \Bigl\{\sum^{\infty}_{n=1} (-1)^{n-1}  
 \Bigl[ \frac{a^{3}}{2 n} K_{1}(n a) + \frac{a^{2}}{n^{2}} K_{2}(n a) + \frac{a^{3}}{2 n} K_{3}(n a) 
 \Bigr]\Bigr\} \; \; , \; \; \label{form9a}
\end{eqnarray}
where $a = \frac{m c^{2}}{T} = \frac{1}{\theta}$ is the inverse temperature (in $keV$). 

\subsection{Ultra-relativistic limit for the relativistic electron-positron plasma}  

Let us discuss the ultra-relativistic limit for the relativistic electron-positron plasma (i.e., $g = 2 s + 
1 = 2$) which is in thermal equilibrium with its annihilation radiation. In such a limit all positrons and 
electrons are considered as some relativistic particles for which $c p \gg m c^{2}$ (or $p \gg m c)$, i.e., 
we can neglect by their energies at rest $E = c p$. The total energy of this plasma is determined by the 
formula  
\begin{eqnarray}
 E_e = E_p = \frac{V}{\pi^{2} \hbar^{3}} \int^{+\infty}_{0} \frac{c p^{3} dp}{\exp\Bigl(\frac{p c}{T}\Bigr)   
 + 1} = \frac{V T^{4}}{\pi^{2} (\hbar c)^{3}} \int^{+\infty}_{0} \frac{x^{3} dx}{\exp(x) + 1} = 
 \frac{7}{8} \; \Gamma(4) \; \zeta(4) \frac{V T^{4}}{\pi^{2} (\hbar c)^{3}} \; \; , \; \; \label{relform1}
\end{eqnarray}
where $x = \frac{c p}{T}$ is our new variable. The total $E_e = E_p$ energies of the ultra-relativistic 
electron-positron plasma is 
\begin{eqnarray}
 E_e = E_p = \frac{7}{8} \; 3! \; \frac{2^{4 - 1} \pi^{4} |B_4|}{4!} \frac{V T^{4}}{\pi^{2} (\hbar c)^{3}} = 
 \frac{7 \pi^{2}}{120} \frac{V T^{4}}{(\hbar c)^{3}} \; \; , \; \;  \label{relform2}
\end{eqnarray}
where $B_4 = -\frac{1}{30}$ is the fourth Bernoulli number. The total number $N_e = N_p$ of electrons and/or 
positrons in the volume $V$ at this temperature $T$ equals
\begin{eqnarray}
 N_e = N_p &=& \frac{V}{\pi^{2} \hbar^{3}} \int^{+\infty}_{0} \frac{c p^{2} dp}{\exp\Bigl(\frac{p c}{T}\Bigr) 
 + 1} = \frac{V T^{3}}{\pi^{2} (\hbar c)^{3}} \int^{+\infty}_{0} \frac{x^{2} dx}{\exp(x) + 1} \nonumber \\ 
 &=& \frac{3}{4} \Gamma(3) \; \zeta(3) \frac{V T^{3}}{\pi^{2} (\hbar c)^{3}} = 1.202056903159594 \; 
 \Bigl(\frac{3}{2 \pi^{2}}\Bigr) \; \frac{V T^{3}}{(\hbar c)^{3}} \; \; , \; \; \label{relform3}
\end{eqnarray}
The last two equations exactly coincide with the results presented in \cite{LLSF}.  
 
\section{Ultra-relativistic Fermi gas} 

As mentioned above the relativistic gas of Fermi particles does not (and cannot) exist without radiation. Indeed, 
collisions between electrically charged fast particles will always produce breaking radiation, or bremsstrahlung. 
At very high temperatures contributions from annihilation $\gamma-$quanta becomes substantial and increase rapidly, 
when the temperature raises. In reality, by considering the confined Fermi gases at high and very high temperatures 
we always have to deal with the stream of thermal and annihilation $\gamma-$quanta and discuss thermal equilibrium 
between these Fermi gases and such a radiation (see, our analysis in the previous Section). In reality, this fact 
substantially simplified our analysis, since we could assume that the sum of chemical potentials of the electron 
and positron gases equal zero identically. Now, we want to consider a pure relativistic gas of Fermi-particles 
which has some positive chemical potential $\mu$. In general, the chemical potential $\mu$ is explicitly included 
in the Fermi-Dirac distribution function. Therefore, in this case we cannot simplify the explicit forms any of 
the arising integrals. However, some limiting cases can be considered analytically, e.g., we can investigate 
some ultra-relativistic gas of fermions. For our analysis below, the following three conditions are crucially 
important and each of them must always be obeyed. First, the chemical potential $\mu$ is different from zero. 
Second, there is no annihilation of fermions into photons. Third, we shall assume that our fermion gas is not in 
thermal equilibrium with the gas of photons. 

Thermal energy $E$ of the relativistic Fermi gas is written in the form 
\begin{eqnarray}
  E = \frac{g V}{2 \pi^{2} \hbar^{3}} \int_{0}^{\infty} \frac{c \sqrt{p^{2} + m^{2} c^{2}} \; p^{2} \;
  dp}{\exp\Bigl(\sqrt{\frac{p^{2} c^{2}}{T^{2}} + \frac{m^{2} c^{4}}{T^{2}}} - 
  \frac{\mu}{T}\Bigr) + 1} = \frac{g V T^{4}}{2 \pi^{2} (\hbar c)^{3}} \int_{0}^{\infty} \frac{\sqrt{y^{2} + 
  a^{2}} \; y^{2} \; dy}{\exp\Bigl(\sqrt{y^{2} + a^{2}} - \frac{\mu}{T}\Bigr) + 1} 
  \; \; , \; \; \label{urelfor1}
\end{eqnarray}  
where $a = \frac{m c}{T} \ge 0 \; , \; a^{2} = \frac{m^{2} c^{2}}{T^{2}}$ and $y = \frac{p c}{T}$, i.e., $dy = 
\frac{T}{c} dp$. This integral can also be represented as an infinite sum, but the final expression is very 
difficult for practical use. Therefore, let us simplify the problem and consider the relativistic Fermi gas in 
the ultra-relativistic limit. This case is, in a certain sense, simpler than the general one and is of 
significant independent interest for numerous applications. In addition to this, in order to solve this 
problem we use a different method. 

In the ultra-relativistic limit we always have $p \gg m c$, and in Eq.(\ref{urelfor1}) it is possible to write 
$\sqrt{p^{2} + m^{2} c^{2}} \approx p$ and $\sqrt{y^{2} + a^{2}} \approx y$. Finally, the integral in the last 
equation, Eq.(\ref{urelfor1}), takes the form     
\begin{eqnarray}
  E = \frac{g V T^{4}}{2 \pi^{2} (\hbar c)^{3}} \int_{0}^{\infty} \frac{y^{3} \; dy}{\exp\Bigl(y - 
  \frac{\mu}{T}\Bigr) + 1} = \frac{g V T^{4}}{2 \pi^{2} (\hbar c)^{3}} \int_{0}^{\infty} \frac{y^{3} \; 
  dy}{\exp\Bigl(y - b\Bigr) + 1} \; \; . \; \; \label{urelfor2} 
\end{eqnarray}  
The integral in this equation does not depend upon the parameter $a = \frac{m c}{T}$ defined above. The only 
parameter of the problem is the ratio of the chemical potential and temperature $b = \frac{\mu}{T} >0$. In this 
notation we can represent the integral in Eq.(\ref{urelfor2}) in the form 
\begin{eqnarray}
  I = \int_{-b}^{\infty} \frac{(b + x)^{3} \; dx}{\exp(x) + 1} = \int_{0}^{b} \frac{(b - x)^{3} \; 
  dx}{1 + \exp(-x)} + \int_{0}^{\infty} \frac{(x + b)^{3} \; dx}{\exp(x) + 1} = I_1(b) + I_2(b)  
  \; \; . \; \; \label{urelfor3} 
\end{eqnarray}  
The second integral in this formula is reduced to the form 
\begin{eqnarray}
  I_2(b) &=& \int_{0}^{\infty} \frac{(x + b)^{3} \; dx}{\exp(x) + 1} = 
  \int_{0}^{\infty} \frac{x^{3} \; dx}{\exp(x) + 1} + 3 b \int_{0}^{\infty} \frac{x^{2} \; dx}{\exp(x) + 1} 
  \nonumber \\
  &+& 3 b^{2} \int_{0}^{\infty} \frac{x \; dx}{\exp(x) + 1} + b^{3} \int_{0}^{\infty} \frac{dx}{\exp(x) + 1}
  \; \; . \; \; \label{urelfor33} 
\end{eqnarray}  
Three of these four integrals are determined by using the following formula (see, the third equation in 
Eq.(3.411) from \cite{GR}) 
\begin{eqnarray}
  \int_{0}^{\infty} \frac{x^{p-1} \; dx}{\exp(q x) + 1} = \frac{1}{q^{p}} \;\Bigl( 1 - \frac{1}{2^{p-1}} 
  \Bigr) \Gamma(p) \zeta(p) \; \; , \; \; \label{urelfor34}  
\end{eqnarray}  
where in our case $q = 1$ and $p = 4, 3, 2$. For $p = 1$ the formula, Eq.(\ref{urelfor33}), cannot be 
applied, but the corresponding (fourth) integral in the right side of Eq.(\ref{urelfor33}) equals $\ln 2$. 
The final form of Eq.(\ref{urelfor33}) is 
\begin{eqnarray}
  I_2(b) &=& \int_{0}^{\infty} \frac{(x + b)^{3} \; dx}{\exp(x) + 1} = 
  \frac{7}{8} \Gamma(4) \zeta(4) + \frac{9}{4} \Gamma(3) \zeta(3) \; b + \frac32 \Gamma(2) \zeta(2) \; 
  b^{2} + (\ln 2) \; b^{3} \nonumber \\
  &=& \frac{7 \pi^{4}}{120} + \Bigl(\frac{9}{2}\Bigr) \; \zeta(3) \; b  + \Bigl(\frac{3 \pi^2}{2}\Bigr) 
  \; b^2 + (\ln 2) \; b^{3} \; \; , \; \; \label{urelfor35} 
\end{eqnarray}  
where $\zeta(3) = 1.202056903159594285399\ldots$. Note that this formula is a finite polynomial of power 
three upon $b = \frac{\mu}{T}$ and this fact crucially simplifies analysis of thermodynamic properties of 
the ultra-relativistic Fermi gas/plasma.  

Now, consider the first integral in Eq.(\ref{urelfor3}). This integral can be written in the form 
\begin{eqnarray}
 I_1(b) = \int_{0}^{b} \frac{(b - x)^{3} \; dx}{1 + \exp(-x)} =  \frac{b^{4}}{4} + \sum_{n=1}^{\infty}
 \frac{(-1)^{n}}{n^{4}} \exp(- b n) \; \gamma(4, -b n) \; \; , \; \; \label{urelfor36}  
\end{eqnarray}  
where $\gamma(a, x)$ is the incomplete $\gamma-$function defined exactly as in \cite{GR}. Since in our case 
the first argument of the $\gamma(a, x)$ function is integer, then we can write 
\begin{eqnarray}
 \gamma(4, -b n) = 3! \Bigl[ 1 - \exp(b n) \sum^{3}_{m=0} \frac{(-b n)^{m}}{m!} \Bigr] = 6 \Bigl[ 1 - \exp(b n) 
 \Bigl( 1 - b n + \frac{b^{2} n^{2}}{2} - \frac{b^{3} n^{3}}{6} \Bigr)\Bigr] \; \; , \; \; \label{urelfor363}  
\end{eqnarray}  
where $b = \frac{\mu}{T}$. These simple analytical formulas for the $I_1(b)$ and $I_2(b)$ integrals completely 
determine the thermal energy $E \simeq I_1(b) + I_2(b)$ of the ultra-relativistic Fermi gas. 

Analogous calculations for the total number of fermions $N$ in the volume $V$ are even simpler. In general, 
for ultra-relativistic Fermi gas this number $N$ is given by the formula 
\begin{eqnarray}
  N = \frac{g V T^{3}}{2 \pi^{2} (\hbar c)^{3}} \int_{0}^{\infty} \frac{y^{2} \; dy}{\exp\Bigl(y - 
  \frac{\mu}{T}\Bigr) + 1} = \frac{g V T^{3}}{2 \pi^{2} (\hbar c)^{3}} \int_{-b}^{\infty} 
  \frac{(x + b)^{3} dx}{\exp(x) + 1} \; \; . \; \; \label{urelfor37} 
\end{eqnarray}  
The integral $J$ in this formula is represented as the sum of the two following integrals $J(b) = J_1(b) 
+ J_2(b)$, where 
\begin{eqnarray}
 J_2(b) &=& \int_{0}^{\infty} \frac{x^{2} \; dx}{\exp(x) + 1} + 2 b \; \int_{0}^{\infty} \frac{x \; dx}{\exp(x) 
 + 1} + b^{2} \; \int_{0}^{\infty} \frac{dx}{\exp(x) + 1} \nonumber \\
 &=& \frac32 \; \zeta(3) + b \; \zeta(2) + b^{2} \; \ln 2 =
 \frac32 \; \zeta(3) + b \; \frac{\pi^{2}}{2} + b^{2} \; \ln 2 \; \; , \; \label{urelfor4a}
\end{eqnarray}  
where $b = \frac{\mu}{T}, \zeta(3) = 1.202056903159594285399\ldots$ (see above) and 
\begin{eqnarray}
 J_1(b) &=& \int_{0}^{b} (b - x)^{2} dx + \int_{0}^{b} (b - x)^{2} dx \Bigl[ \sum^{\infty}_{n=1} \exp(-n x) 
 \Bigr] \nonumber \\
  &=& \frac56 \; b^{3} + \sum_{n=1}^{\infty} \frac{(-1)^{n-1}}{n^{3}} \exp(- b n) \; \gamma(3, -b n) \; 
  \; , \; \; \label{urelfor4b}  
\end{eqnarray}  
where $\gamma(a, x)$ is the incomplete $\gamma-$function mentioned above. In our present case $a = 3$ and we 
can write    
\begin{eqnarray}
 \gamma(3, -b n) = \gamma(1 + 2, -b n) = 2 \Bigl[ 1 - \exp(n b) \Bigl(1 - b n + \frac{b^{2} n^{2}}{2}\Bigr) 
 \Bigr] \; \; , \; \label{urelfor5}   
\end{eqnarray}  
where again $b = \frac{\mu}{T}$. 

Now, it easy to derive the final formulas for the energy $E$ and total number of fermions $N$ in the 
ultra-relativistic Fermi gas. Thermal energy $E$ of this gas is  
\begin{eqnarray}
 E(b) = \frac{g V T^{4}}{2 \pi^{2} (\hbar c)^{3}} \Bigl[\frac{7 \pi^{4}}{120} &+& \Bigl(\frac{9}{2}\Bigr) 
 \; \zeta(3) \; b + \Bigl(\frac{3 \pi^2}{2}\Bigr) \; b^2 + (\ln 2) \; b^{3} + \frac{b^{4}}{4} \nonumber \\
 &+& \sum_{n=1}^{\infty} \frac{(-1)^{n}}{n^{4}} \exp(- b n) \; \gamma(4, -b n) \Bigr] = V T^{4} f_{1} 
 \Bigl(\frac{\mu}{T}\Bigr) \; \; , \; \; \label{urelfor5aa} 
\end{eqnarray}   
where $b = \frac{\mu}{T}$ and the scalar function of one variable $f_{1}$ is 
\begin{eqnarray}
  f_1(b) = \frac{g}{2 \pi^{2} (\hbar c)^{3}} \Bigl[\frac{7 \pi^{4}}{120} &+& \Bigl(\frac{9}{2}\Bigr) 
  \; \zeta(3) \; b + \Bigl(\frac{3 \pi^2}{2}\Bigr) \; b^2 + (\ln 2) \; b^{3}  + \frac{b^{4}}{4} 
  \nonumber \\
  &+& \sum_{n=1}^{\infty} \frac{(-1)^{n}}{n^{4}} \exp(- b n) \; \gamma(4, -b n) \Bigr] \label{urelfor5a0} 
\end{eqnarray} 
For the total number of fermions $N$ we can now write the following formula  
\begin{eqnarray}
 N = \frac{g V T^{3}}{2 \pi^{2} (\hbar c)^{3}} \Bigl[\frac32 \; \zeta(3) &+& b \; \frac{\pi^{2}}{2} + b^{2} 
 \; \ln 2 + \frac56 \; b^{3} \nonumber \\
 &+& \sum_{n=1}^{\infty} \frac{(-1)^{n-1}}{n^{3}} \exp(- b n) \; \gamma(3, -b n) 
 \Bigr] = V T^{3} f_{0}\Bigl(\frac{\mu}{T}\Bigr)\; \; , \; \; \label{urelfor5bb}
\end{eqnarray} 
where 
\begin{eqnarray}
 f_0(b) = \frac{g}{2 \pi^{2} (\hbar c)^{3}} \Bigl[\frac32 \; \zeta(3) + b \; \frac{\pi^{2}}{2} + b^{2} \; 
 \ln 2 + \frac56 \; b^{3} + \sum_{n=1}^{\infty} \frac{(-1)^{n-1}}{n^{3}} \exp(- b n) \; \gamma(3, -b n) 
 \Bigr] \; \; . \; \label{urelfor5b0}
\end{eqnarray} 
The equation of state for such an ultra-relativistic Fermi gas is written in the form $\Omega = - \frac13 E$, 
or $P = T^{4} f_1\Bigl(\frac{\mu}{T}\Bigr)$, where $f_1(b)$ is the real function of one variable only. This 
function is defined in Eq.(\ref{urelfor5a0}) above. 

\section{Conclusion} 

We have considered thermodynamic properties of the electron-positron plasma (or gas) at high and very high 
temperatures. By using our approach we have derived the explicit formulas which can be used for numerical 
computations of basic thermodynamic properties of arbitrary, in principle, Fermi gases with the both positive 
and negative chemical potentials. This our approach works well for the high-temperature, gravitationally 
confined plasma which has a unique ability to generate electron-positron pairs in very large numbers (for $T 
\ge$ 170 $keV$). The arising $(e^{-}, e^{+})-$pairs also annihilate into a few $\gamma-$quanta. At similar 
conditions the total numbers of newly created electrons and positrons (per unit volume) significantly exceed 
the total number of initial particles in the same volume, i.e., atomic electrons and nuclei. Thus, in the 
result of high-temperature heating of some confined, relatively dense atomic plasma one always will end up 
with the electron-positron plasma. The density of such an electron-positron plasma rapidly increases with the 
temperature $\rho_{e^{-}, e^{+}} \simeq T^{4}$, while the role of incident particles in similar confined, 
high-temperature plasma with $T \ge$ 170 $keV$ becomes negligible. For higher temperatures, e.g., for $T \ge$ 
350 $keV$, any heated and confined plasma, which is relatively dense, will essentially consist of electrons 
and positrons only. In general, such a plasma will emit extremely large numbers of annihilation $\gamma-$quanta 
with possible admixture of fast positrons and electrons.


\begin{thebibliography}{01}

\bibitem{Fro1} A.M. Frolov, \textit{Bound state properties and positron annihilation in the negatively 
charged Ps$^{-}$ ion. On thermal sources of annihilation $\gamma$-quanta in our Galaxy}, ArXiv: 4682677 
[phys.atom-ph.] (8th of January 2023). 

\bibitem{Planck} M. Planck, \textit{Theory of Heat Radiation.} (Dover, New York (1959)). 

\bibitem{BornW} M. Born and E. Wolf, \textit{Principles of Optics.} (4th edition, Pergamon Press, New York (1968)). 

\bibitem{Panth1} F.H. Panther, R.M. Crocker, Y. Birnboim, I.R. Seitenzahl and A.J. Ruiter, Monthly Notices 
Royal Astron. Soc. {\bf 474}, L17 (2018). 

\bibitem{Panth2} F.H. Panther, Galaxies {\bf 6}, 39 (2018). 

\bibitem{LLSF} L.D. Landau and E.M. Lifshitz, \textit{Statistical Physics. Course of Theoretical Physics. 
Volume 5} (3rd edition, Butterworth-Heinemann, Oxford, UK (1980)).  

\bibitem{Feynman} R.P. Feynman, \textit{Statistical Mechanics. A set of Lectures} (W.A. Benjamin, Inc., Boston, MA 
(1972)).  

\bibitem{Mayer} J.E. Mayer and M. Goeppert Mayer, \textit{Statistical Mechanics} (2nd edition, John Willey \& Sons, 
Inc., New York (1977)). 

\bibitem{EHSE} H. Eyring, D. Henderson, B.J. Stover and E.M. Eyring, \textit{Statistical Mechanics and Dynamics} 
(J. Wiley \& Sons Inc., New York (1964)). 

\bibitem{XXX} In general, the relation between $\Omega-$potential (or the product $P V$) and energy $E$ is called 
the equation of state and it is written in the form $\Omega = \Omega(E)$. The equation of state plays a crucial role 
in thermodynamics of actual gases, including Fermi and Bose gases and/or plasmas. 

For all Fermi systems considered in this
study the equations of state are reduced to the form $P = P(T; \mu)$.    

\bibitem{GR} I.S. Gradstein and I.M. Ryzhik, \textit{Tables of Integrals, Series and Products} (6th revised ed., 
Academic Press, New York (2000)). 

\bibitem{Mac} H.M. Macdonald, Proc. London Math. Soc. {\bf XXX}, 167 (1899). 

\bibitem{Watson} G.N. Watson, \textit{a treatise on the THEORY OF BESSEL FUNCTIONS.} 
(2nd edition, Cambridge University Press, London (1944), reprinted in 1966). 

\bibitem{Teul} W.H. Press, S.A. Teulkolsky, W.T. Vetterling and B.P. Flannery, {\it Numerical Recipes in 
Fortran 77. The art of Scientific Computing}, (2nd. ed., Canbridge University Press, Cambridge, UK (1996)), Chpt. 10.

\bibitem{NIST} see, e.g., https://physics.nist.gov/cgi-bin/cuu/Value?

\end{thebibliography}
\end{document}